\documentclass[epj]{svjour}
\usepackage{graphics}
\begin{document}
\title{
Analyzing powers  $A_{yy}$, $A_{xx}$, $A_{xz}$ 
and $A_{y}$ in the $dd \to  {\rm ^3He}n$
\\
reaction at  270 MeV 
}


\author{
 M.~Janek\inst{1,2}\thanks{\emph{E-mail address:} janek@sunhe.jinr.ru} \and
 T.~Saito\inst{3}     \and
V.P.~Ladygin\inst{1}\thanks{\emph{E-mail address:} ladygin@sunhe.jinr.ru} \and 
T.~Uesaka\inst{4}\thanks{\emph{E-mail address:} uesaka@cns.s.u-tokyo.ac.jp}    \and
M.~Hatano\inst{3}    \and 
A.Yu.~Isupov\inst{1} \and
H.~Kato\inst{3}      \and
N.B.~Ladygina\inst{1}\and 
Y.~Maeda\inst{4}     \and
A.I.~Malakhov\inst{1}\and 
J.~Nishikawa\inst{5} \and
T.~Ohnishi\inst{6}       \and
H.~Okamura\inst{7}       \and
S.G.~Reznikov\inst{1} \and  
H.~Sakai\inst{3,4}    \and  
S.~Sakoda\inst{3}     \and
N.~Sakamoto\inst{6}   \and 
Y.~Satou\inst{8}      \and
K.~Sekiguchi\inst{6}  \and
K.~Suda\inst{4}       \and
A.~Tamii\inst{8}       \and
N.~Uchigashima\inst{3} \and 
T.A.~Vasiliev\inst{1}  \and
K.~Yako\inst{3}     
 }                     
%

  \institute{
VBLHE-JINR, 141980 Dubna, Moscow region, Russia \and
University of P.J. \v{S}af\'arik, 041-54 Ko\v{s}ice, Slovakia \and
Department of Physics, University of Tokyo, Bunkyo, Tokyo 113-0033, Japan \and
Center for Nuclear Study, University of Tokyo, Bunkyo, Tokyo 113-0033, Japan \and
Department of Physics, Saitama University, Urawa 338-8570, Japan \and
RIKEN, Wako, Saitama 351-0198, Japan \and
CYRIC, Tohoku University, Sendai, Miyagi 980-8578, Japan \and
Research Center for Nuclear Physics, Osaka University, Ibaraki 567-0047, Japan 
  }

  \date{Received: date / Revised version: date}
%

  \abstract{
The data on the tensor $A_{yy}$, $A_{xx}$, $A_{xz}$ and vector $A_y$ analyzing powers in the $dd \to {\rm ^3He}n$
	obtained  at $T_d$~= 270~MeV in the angular range 0$^\circ$ -- 110$^\circ$ in the c.m.  are presented.  
The observed negative sign of the tensor analyzing powers 
	$A_{yy}$, $A_{xx}$ and $A_{xz}$ at small angles clearly demonstrate the
	sensitivity to the	ratio of the $D$ and $S$ wave component of the ${\rm ^3He}$ wave function. 
However, the one-nucleon exchange calculations by using
	the standard ${\rm ^3He}$ wave functions have failed to reproduce 
	the strong variation of the tensor analyzing powers as a function of the angle in the c.m.    
    } 
  \maketitle
\section{Introduction}
\label{introduction}

Intensive theoretical and experimental efforts performed 
	during last years led to a new generation of realistic nucleon-nucleon ($NN$)
	potentials like  AV-18 \cite{kim3}, CD-Bonn \cite{kim4}, Nijmegen I, II and 93 \cite{kim7} etc. 
These potentials describe the existing $NN$ scattering data up to 350 MeV with an
	unprecedented precision.
However,  already in the elastic $Nd$ scattering there are significant discrepancies
	between the measured observables and the Faddeev calculations based on
	pairwise $NN$ potentials~(see review~\cite{glock} and references therein).
A part of this  discrepancy in the cross section at the energies $\le$135 MeV/nucleon
	\cite{kimiko0,ermish0} has been reduced by including three nucleon forces ($3NF$). 
At higher kinetic energies, the backward angles require 
	more sophisticated approaches with a new type of $3NF$ and/or relativistic corrections  
	\cite{ermish0,hatanaka}.   
On the other hand,  Faddeev calculations cannot reproduce
	the behavior of the polarization observables in
	the $dp$ elastic scattering \cite{hatanaka,sak1,kimiko1,pd_ay,cadman,kimiko2,ermish}. 
These results clearly indicate  deficiencies in the spin-dependent part of the 3NF models used in the calculations.

In this respect, three nucleon  bound states are of particular interest, because even such  
	a fundamental quantity as the binding energy of the system 
	cannot be reproduced  by calculations with
	modern pairwise nucleon-nucleon potentials~\cite{glock}.
Since the binding energy is known to be closely related with the power 
	of spin-dependent forces such as the tensor  and/or three-nucleon forces,
	an experimental study of the 
	spin structure of three-nucleon bound system is crucial  to understand the source of  underbinding.

The non-relativistic Faddeev calculations~\cite{theory}  for three-nucleon 
	bound state have predicted  that the dominant components 
	of the ${\rm ^3He}$  ground state are as follows: a spatially symmetric $S$-state, 
	where the ${\rm ^3He}$ spin due to the neutron  and 
	two protons are in a spin singlet state; 
	and a $D$-state, where all three nucleon spins are oriented opposite to the ${\rm ^3He}$ spin.
The $S$-state is found to dominate at small momenta while $D$-state dominates at large momenta. 
The relative sign of the $D$- and $S$- waves in the momentum space is positive at small 
	and moderate nucleon momenta \cite{santos}.
The data sensitive to the three-nucleon bound state spin structure 
	are 	scarce and new polarization data, especially at short internucleonic distances, are of great importance.

The ${\rm ^3He}$ structure information is contained in the spin-dependent spectral
	function  $S_{\hat\sigma}^N(E,{\bf q})$ \cite{mil4}, defined as the probability density of  nucleon $N$ 
	found in the  ${\rm ^3He}$ nucleus with  separation energy $E$, momentum ${\bf q}$, 
	and spin along (opposite to)
	the ${\rm ^3He}$ spin indicated by ${\hat\sigma}$=+(--).
The nucleon momentum distribution in ${\rm ^3He}$ is described by the 
	spin-averaged spectral function
	$S^N(E,{\bf q})$.

\begin{sloppypar}
The nucleon momentum distribution, or $S^N(E,{\bf q})$, was investigated by the reactions 
	of quasielastic knockout of the ${\rm ^3He}$ constituent nucleons. The spectral functions
	$S^N(E,{\bf q})$ extracted by the plane-wave impulse approximation (PWIA) 
	analysis from the ${\rm ^3He}(e, ep)$ \cite{jan5}, 
	${\rm ^3He}(p, 2p)d$, and ${\rm ^3He}(p, pd)p$ reactions \cite{eps6}, were found to be in  
	a reasonable agreement.

To investigate the spin-dependent spectral function $S_{\hat\sigma}^N(E,{\bf q})$,
	  one needs to measure the polarization observables. 
Spin correlations for the quasielastic ${\rm ^3He}(p, pN)$ reactions were measured 
	at IUCF Cooler Ring \cite{mil7} up to the internal nucleon momentum q $\sim$ 400 MeV/$c$. 
The spin asymmetry in the momentum distribution
	 proportional to $S_{+}^N(E,{\bf q})$ - $S_{-}^N(E,{\bf q})$, 
	was extracted from the experimental results by the PWIA analysis and compared with the Faddeev solution. 
A good agreement of the experimental neutron and proton spin-dependent spectral functions  
	with the  Faddeev calculations  \cite{theory}
	was observed  at low nucleon momenta. However,
	there is a discrepancy between the experiment and theory in the region
	of q $\ge$ 300 MeV/$c$. 
This deviation can be caused by the uncertainty of  the high-momentum spin structure of the ${\rm ^3He}$  
	as well as by the reaction mechanisms which have not been taken into account in  PWIA.

The radiative deuteron-proton capture reaction, $dp\to{\rm ^3He}\gamma$, 
	at intermediate energies involves a large momentum transfer and therefore can be used to study  
	high momentum components of the ${\rm ^3He}$ wave function. 
The measurements of the tensor analyzing powers \cite{capture1} have shown their sensitivity to the $D$-state 
	component in ${\rm ^3He}$.
Recently the vector $A_y$ and tensor $A_{yy}$, $A_{xx}$ analyzing powers have been measured at $KVI$
	at 55, 66.5 and  90~MeV/nucleon \cite{meh07}. 
The data are in a good agreement with the  results of Faddeev calculations 
	obtained by the Bochum-Cracow  \cite{BC17} and Hannover groups \cite{han19},
	which have shown the effect of $3NF$ to be small at these energies.  
However,  the $KVI$ data \cite{meh07} are in strong contradiction
	with the $A_{xx}$ data obtained at 100 MeV/nucleon at $RCNP$ \cite{sag03}.

The $d{\rm ^3He}$- backward elastic scattering  and $dd\to {\rm ^3He}n({\rm ^3H}p)$ reactions at intermediate 
	and high energies are the one-nucleon exchange (ONE)  processes with a large momentum
	transfer and, therefore, can be used as an effective tool to investigate the
	${\rm ^3He}$ structure at short distances.  
The theoretical analysis of the $dd\to {\rm ^3He}n$ reaction \cite{lad} 
	performed within ONE approximation, has shown that the tensor analyzing powers  due to
	polarization of the incident deuteron are sensitive to the spin distribution of neutron, 
	when ${\rm ^3He}$ is emitted in the forward direction in the c.m.  
Tensor analyzing powers are sensitive to the ratio 
	of  the $D$ and $S$ wave component of the ${\rm ^3He}$ wave function 
	at intermediate energies \cite{lad,r308n}.

The polarization data for $dd$- scattering at intermediate energies are scarce. 
But more data have been obtained on the analyzing powers in the $dd$- elastic scattering \cite{ddelas_low},
	$dd\to{\rm ^3He}n$ and $dd\to{\rm ^3H}p$ reactions \cite{ddhe3nlow}	at low energies. 
These data have been reproduced with the four-body calculations 
	by solving Alt-Grassberger-Sandhas  \cite{fonseca} and  Faddeev-Yakubovsky \cite{carbonell} equations,
	and hyperspherical harmonics methods \cite{kievsky}.
However, at the moment these calculations  cannot be applied to the higher energies.

The experiment  on the  measurement of  the energy and angular 
	dependences of the tensor analyzing powers in the $dd\to {\rm^3He}n({\rm ^3H}p)$ process 
	in the conditions, when the contribution from the $D$- state in ${\rm ^3He(^3H)}$ becomes larger,
has been performed at RIKEN. 
Obtained at  the deuteron kinetic energy of  140, 200 and 270 MeV the data on the tensor analyzing powers $T_{20}$ 
	in the $dd\to {\rm ^3He(0^\circ)}n({\rm ^3H(0^\circ)}p)$  reaction  have positive values \cite{l2004}, 
	which are in a good agreement with the $T_{20}$  data in $d{\rm ^3He}$- backward elastic scattering  \cite{dhe3} 
	and clearly demonstrate the sensitivity to the  $D$- wave effect in the three nucleon bound states.

This paper gives the data on the angular distribution of the analyzing powers 
	$A_{yy}$, $A_{xx}$, $A_{xz}$ and $A_y$  in the $dd\to {\rm ^3He}n$  reaction
	at 270~MeV of the deuteron kinetic energy.
The details of the experimental procedure are described in section~\ref{experiment},
	the results are discussed in section~\ref{results}, the conclusions are written in section~\ref{conclusions}.

\end{sloppypar}

\section{Experiment}
\label{experiment}

The experiment has been  performed at RIKEN Accelerator Research Facility.
The details of the experiment were discussed elsewhere \cite{l2004}, below we mention briefly
	the main items of the experimental procedure.

The high-intensity polarized deuteron beam was produced by the polarized ion source (PIS) \cite{source} 
	and accelerated  by the AVF and Ring Cyclotrons up to the energy of 270 MeV. 
The direction of the symmetry axis of the beam polarization was controlled with a Wien-filter located at 
	the exit of the PIS.

\begin{figure}
\resizebox{0.48\textwidth}{!}{
  \includegraphics{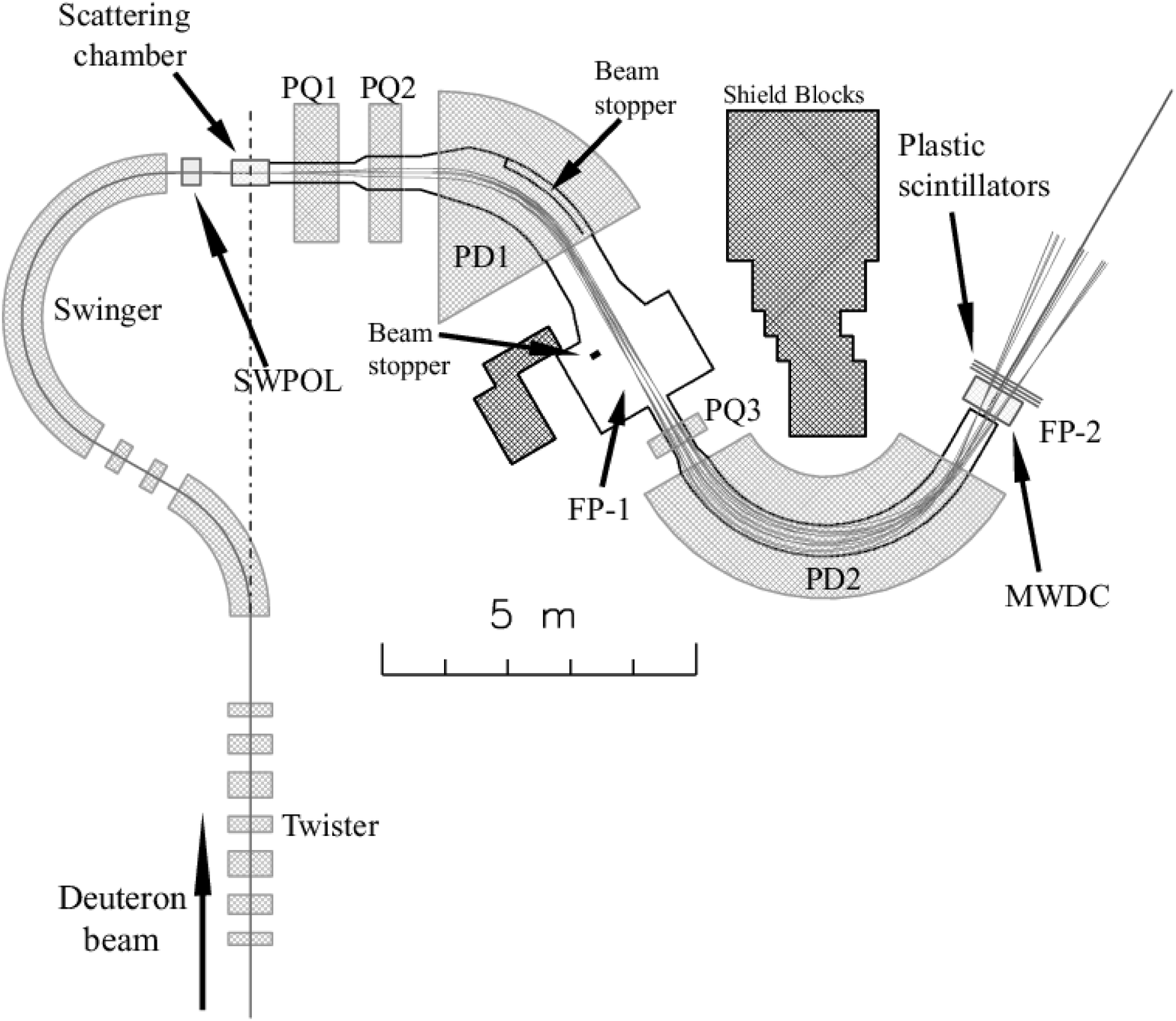}
}
\caption{Spectrometer SMART and the detection system.
$PD_i$ and $PQ_i$ are the dipole and quadrupole magnets, respectively;
$MWDC$ are the multiwire drift chambers; $SWPOL$ is the beam-line deuteron polarimeter;
$FP1$ and $FP2$ are the focal planes of the spectrometer.} 
\label{smart}
\end{figure}

\begin{sloppypar}

The polarization of the deuteron beam was measured with two beam-line polarimeters. 
The $dp$ elastic scattering was used with the known large values of the tensor and vector analyzing powers 
	$A_y$, $A_{yy}$, $A_{xx}$ and $A_{xz}$ \cite{sak1,kimiko1}.
These data   of the analyzing powers \cite{sak1,kimiko1} were taken 
	to analyze  ${\rm dp}$- elastic scattering 
	at 140 and 270 MeV.
The values were obtained for the polarized deuteron beam, whose absolute
	polarization  had been calibrated via the ${\rm ^{12}C}(d,\alpha){\rm	^{10}B^*[2^+]}$ reaction \cite{suda}.

The first polarimeter situated downstream of the Ring Cyclotron was 
	used for the beam polarization monitoring while taking data.
The second polarimeter ($SWPOL$) located 
	in front of the scattering chamber in the experimental hall 
	measured  polarization before and after  each run.
The  polarization values obtained from the both polarimeters  agreed 
	with each other within the statistical accuracy, therefore, 
	the beam polarization for each polarization state of the PIS was taken as
	a weighted average of the values obtained by these polarimeters.

In the present experiment the data were taken for the vector and tensor polarization modes 
	which had the following theoretical maximum polarization: 
	 $(p_z,p_{zz})$ =  $(0,0)$, $(0,-2)$, $(-2/3,0)$ and $(1/3,1)$. 
The actual   values of the beam polarization were between 45 and 85 $\%$
 	of the maximum theoretical value.
The systematic error due to the uncertainties of the 
	$dp$-elastic scattering analyzing power values \cite{sak1,kimiko1}
	does not exceed $\sim2\%$ both for the vector and tensor polarization of the beam.
The systematic and statistical errors have been added in quadrature
	 to calculate the total error of the beam polarization values.
\end{sloppypar}

\begin{sloppypar}
The layout of the experiment is shown in Fig.\ref{smart}. 
SMART spectrograph (Swinger and Magnetic Analyzer with a Rotator and a Twister)
	  \cite{smart} was used for these measurements.
The measurements of the particle momentum and separation from the
	primary beam were performed by the magnetic system 
	of SMART spectrograph consisting of two dipole and three quadrupole magnets (Q-Q-D-Q-D).

Two deuterated polyethylene (CD$_2$) sheets of  54~mg/cm$^2$ and 32~mg/cm$^2$
	thick \cite{maeda} placed in the
	scattering chamber of the SMART were taken as the deuterium targets. 
The carbon foil 34 mg/cm$^2$ thick
	was used to measure the background spectra.

The detection system of SMART at the focal point $FP2$  consisted of 
	three plastic scintillation counters and a multiwire drift chamber (MWDC).
The coincidence  of the signal outputs of all the three
	scintillation counters was employed as the event trigger. 
Pulse heights of the plastic scintillation
	counters were used to select the particle of interest at the trigger level.
Protons and deuterons were partly suppressed by raising threshold levels
	of the constant fraction discriminators. 
The fraction of the event rate for single-charged particles   was $\sim40\%$. 
The CFD thresholds were tuned in such a way not to lose the ${\rm ^3He}$ events
	keeping the dead-time of the data acquisition system at the level of 20-30$\%$.
The admixtures of the background events were almost completely
	eliminated by a software cut in the offline analysis.

\begin{figure}
\resizebox{0.48\textwidth}{!}{
  \includegraphics{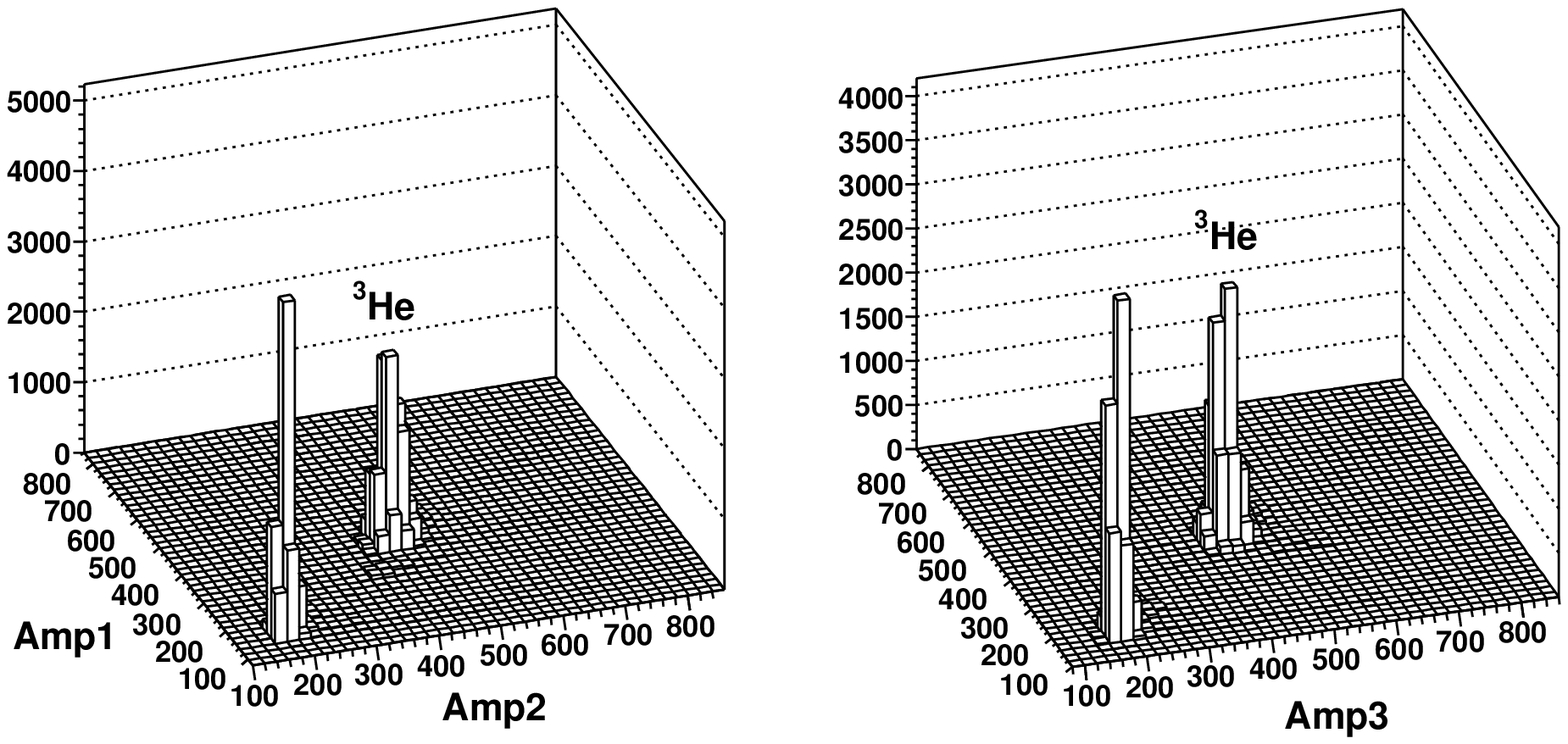}
}
\caption{Correlation of the amplitudes in the scintillation detectors.
The left (right) panel corresponds to the correlation for the
$1^{st}$ and  $2^{nd}$  (the $1^{st}$ and   $3^{rd}$) counters.
}
\label{amplit}

\resizebox{0.48\textwidth}{!}{
  \includegraphics{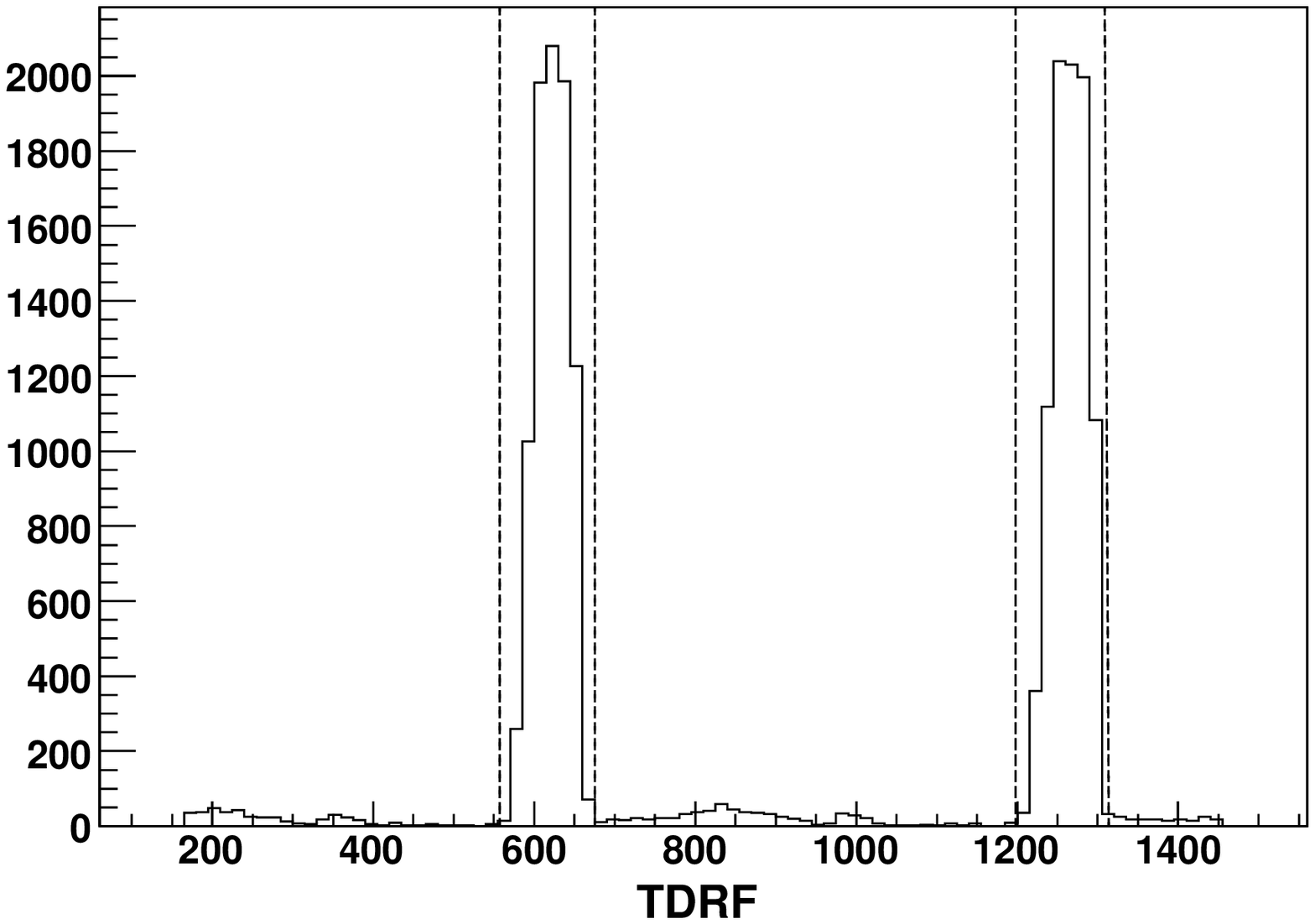}
}
\caption{ The time difference between the trigger signal and the radio-frequency
	signal of the cyclotron. 
The dashed lines represent the imposed windows  to select  the ${\rm ^3He}$ nuclei.
}
\label{tdrf}
\end{figure}

The particle identification was carried out on the energy losses in the plastic 
	scintillators and time-of-flight between the target and the detection point. 
The event was considered as a certain type of particle (${\rm ^3He}$) only
	in the case when  the pulse height was correlated in all three scintillation counters.
The correlation of the amplitudes in the $1^{st}$ and  $2^{nd}$  (the $1^{st}$ and  $3^{rd}$)
	scintillation detectors is presented in the left (right) panel of  Fig.\ref{amplit}.
The distance between the target and the detection point was about 17 m, 
	which was enough to separate ${\rm ^3He}$, deuterons, and protons with the
	same momentum from the time-of-flight (TOF - the time difference between 
	the trigger signal and the radio-frequency signal of the cyclotron, see Fig.\ref{tdrf}).
The start signal to measure  TOF came from the event trigger. 
The imposed windows to select the ${\rm ^3He}$ nuclei are shown 
	in Fig.\ref{tdrf} by the dashed  lines.
 
\end{sloppypar}

\begin{sloppypar}

The MWDC information was taken to reconstruct the particle trajectories in the focal plane $FP2$. 
The trajectories of the detected particles at the second focal plane were determined 
	by the least square method using the position information obtained from the MWDC. 
The typical track reconstruction efficiency of the MWDC was better than 99$\%$. 
The ion-optical parameters of the  SMART spectrograph were also taken into account to calculate
	 the momentum of the particle and emission angle in the target  to obtain the track information. 
The resulting energy resolution was $\sim 300$~keV.

\end{sloppypar}

\begin{figure}
\resizebox{0.48\textwidth}{!}{
  \includegraphics{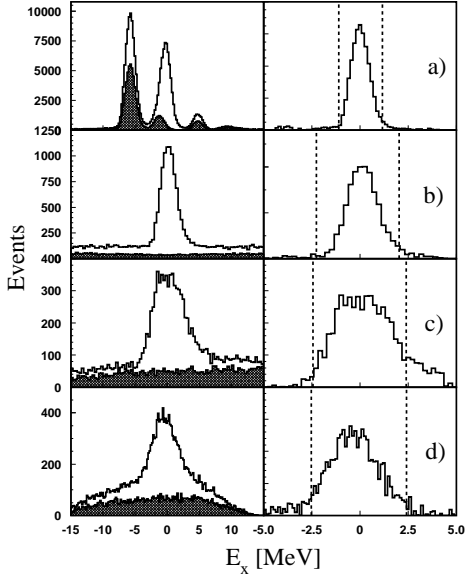}
}
\caption{CD$_2$-C subtraction for the $dd\to {\rm^3He}n$ 
reaction at $T_d$ = 270 MeV. The open and shadowed histograms in
the left panels correspond to the yields from the CD$_2$ and carbon targets, respectively.
The right panels demonstrate the quality of the CD$_2$-C subtraction.
The panels a), b), c) and d) correspond to the ${\rm ^3He}$
scattering angle in the c.m. of 5$^\circ$, 32$^\circ$, 54$^\circ$ and 94$^\circ$, respectively.   
}
\label{vycitanie}
\end{figure}

\begin{sloppypar}

The contribution of the  deuterium target was obtained via CD$_2-$C subtraction procedure
	for each spin state at every angle.
Subtraction procedure is  shown in Fig.\ref{vycitanie} a), b), c) and d)  for the ${\rm ^3He}$
	scattering angle in the c.m. of 5$^\circ$, 32$^\circ$, 54$^\circ$ 
	and 94$^\circ$, respectively.
The spectra are plotted as a function of the excitation energy $E_X$ defined as follows:
\begin{eqnarray}
E_X=\sqrt{(E_0-E_{3N})^2-({\bf P}_0-{\bf P}_{3N})^2}-M_{N},
\end{eqnarray}
	where {\bf P$_0$} is the incident momentum; $E_0=2M_d+T_d$ is the total initial energy;  
	$E_{3N}$ and {\bf P$_{3N}$}  are the energy and momentum 
	of  the three-nucleon system, respectively; $M_{N}$ is the nucleon mass.
The left panels represent the relative yields from the CD$_2$ and carbon targets
	shown by the open and shadowed histograms, respectively. 
The histograms are not normalized for the sake of easy comparison.
Peaks at $E_{X}=0$~MeV correspond to ${\rm ^3He}$ from the $dd\to {\rm ^3He}n$ reaction. 
The right panels show the spectra after subtraction
	of the carbon events normalized on luminosity 
	and  dead-time correction. 
It is clearly demonstrated that the subtraction procedure has been carried out properly.

\end{sloppypar}

The analyzing powers $A_y$, $A_{yy}$, $A_{xx}$ and $A_{xz}$ in the $dd\to{\rm ^3He}n$ reaction 
	were obtained from the number of the events  after the CD$_2-$C subtraction  procedure 
	and beam polarization.
The number of the events was normalized on the dead-time effect, 
	the detection efficiency, and  beam intensity.
When the ${\rm ^3He}$ scattering angle in the c.m. was equal to or less than 7$^\circ$, 
  	the azimuthal angle to cover  the scattered particles became larger.  
In this case the range of the azimuthal angle was divided into bins of 15 degrees.
The asymmetry from each bin for each polarized spin mode of PIS was acquired individually and 
	the analyzing powers were obtained from the fit of the asymmetries distribution by 
	the functions depending on the azimuthal angle.
Since the polarization modes were cycled every 5 seconds, the
	systematic uncertainty due to any time-dependent effects such as
	deuterium loss from the CD$_2$ target caused by  beam irradiation, can be
	neglected.

\section{Results and discussion}
\label{results}

\begin{table}
\begin{minipage}[h]{75mm}
{
\caption{The angular dependence of the vector  analyzing power $A_y$ in the 
$dd\to {\rm ^3He}n$ reaction at 270 MeV.}
}  
{\centering 
\begin{tabular}{llll}
\hline\noalign{\smallskip}
$\theta_{c.m.}$ &   ~~~$A_y\pm\Delta A_y$    & $\theta_{c.m.}$ &   ~~~$A_y\pm\Delta A_y$\\
\noalign{\smallskip}\hline\noalign{\smallskip} 
  1.0  & $-$0.012 $\pm$ 0.026  &   54.0  & $-$0.393 $\pm$ 0.029 \\
  3.0  & $-$0.037 $\pm$ 0.016  &   56.0  & $-$0.326 $\pm$ 0.032 \\
  5.0  & $-$0.019 $\pm$ 0.014  &   58.0  & $-$0.274 $\pm$ 0.024 \\
  7.0  & $-$0.081 $\pm$ 0.012  &   60.0  & $-$0.327 $\pm$ 0.029 \\
 10.0  & $-$0.098 $\pm$ 0.012  &   62.0  & $-$0.348 $\pm$ 0.024 \\
 12.0  & $-$0.136 $\pm$ 0.013  &   64.0  & $-$0.266 $\pm$ 0.024 \\
 14.0  & $-$0.133 $\pm$ 0.015  &   66.0  & $-$0.348 $\pm$ 0.024 \\
 16.0  & $-$0.146 $\pm$ 0.016  &   68.0  & $-$0.326 $\pm$ 0.025 \\
 18.0  & $-$0.129 $\pm$ 0.018  &   70.0  & $-$0.299 $\pm$ 0.024 \\
 20.0  & $-$0.119 $\pm$ 0.022  &   72.0  & $-$0.331 $\pm$ 0.026 \\
 22.0  & $-$0.085 $\pm$ 0.024  &   74.0  & $-$0.271 $\pm$ 0.024 \\
 24.0  & $-$0.075 $\pm$ 0.031  &   76.0  & $-$0.323 $\pm$ 0.027 \\
 26.0  & $-$0.018 $\pm$ 0.016  &   78.0  & $-$0.333 $\pm$ 0.037 \\
 28.0  & $-$0.026 $\pm$ 0.012  &   80.0  & $-$0.290 $\pm$ 0.035 \\
 30.0  & $-$0.023 $\pm$ 0.012  &   82.0  & $-$0.351 $\pm$ 0.033 \\
 32.0  & $-$0.043 $\pm$ 0.012  &   84.0  & $-$0.310 $\pm$ 0.025 \\
 34.0  & $-$0.040 $\pm$ 0.013  &   86.0  & $-$0.172 $\pm$ 0.044 \\
 36.0  & $-$0.075 $\pm$ 0.015  &   88.0  & $-$0.251 $\pm$ 0.044 \\
 38.0  & $-$0.093 $\pm$ 0.013  &   90.0  & $-$0.292 $\pm$ 0.030 \\
 40.0  & $-$0.153 $\pm$ 0.020  &   92.0  & $-$0.349 $\pm$ 0.026 \\
 42.0  & $-$0.166 $\pm$ 0.027  &   94.0  & $-$0.310 $\pm$ 0.024 \\
 44.0  & $-$0.177 $\pm$ 0.029  &   96.0  & $-$0.095 $\pm$ 0.058 \\
 46.0  & $-$0.194 $\pm$ 0.031  &   98.0  & $-$0.181 $\pm$ 0.048 \\
 48.0  & $-$0.282 $\pm$ 0.024  &  100.0  & $-$0.116 $\pm$ 0.062 \\
 50.0  & $-$0.329 $\pm$ 0.030  &  104.0  & $-$0.123 $\pm$ 0.054 \\
 52.0  & $-$0.331 $\pm$ 0.032  &         &                    \\
\noalign{\smallskip}\hline
\end{tabular}
}
\end{minipage}
\label{table_ay}
\end{table}

The results on the angular distribution of the analyzing powers $A_y$, $A_{yy}$, $A_{xx}$ and $A_{xz}$ 
	in the $dd\to {\rm ^3He}n$ reaction at the incident deuteron energy $T_d$= 270 MeV 
	are given in tables 1, 2, 3 and 4, respectively.  
The systematic and the statistical error of  analyzing powers  have been added in quadrature.

\begin{table}
\begin{minipage}[h]{75mm}
{
\caption{The angular dependence of the tensor  analyzing power $A_{yy}$ in the 
$dd\to {\rm ^3He}n$ reaction at 270 MeV.}
}  
{\centering 
\begin{tabular}{llll}
\hline\noalign{\smallskip}
$\theta_{c.m.}$ &  ~~$A_{yy}\pm\Delta A_{yy}$    & $\theta_{c.m.}$ &   ~~$A_{yy}\pm\Delta A_{yy}$       \\
\noalign{\smallskip}\hline\noalign{\smallskip} 
  1.0  & $-$0.184 $\pm$ 0.027 &   54.0  & $-$0.139 $\pm$ 0.041 \\
  3.0  & $-$0.204 $\pm$ 0.018 &   56.0  & $-$0.085 $\pm$ 0.044 \\
  5.0  & $-$0.209 $\pm$ 0.027 &   58.0  &  0.015 $\pm$ 0.032 \\
  7.0  & $-$0.192 $\pm$ 0.033 &   60.0  &  0.047 $\pm$ 0.040 \\
 10.0  & $-$0.155 $\pm$ 0.017 &   62.0  &  0.159 $\pm$ 0.031 \\
 12.0  & $-$0.116 $\pm$ 0.018 &   64.0  &  0.173 $\pm$ 0.031 \\
 14.0  & $-$0.027 $\pm$ 0.019 &   66.0  &  0.198 $\pm$ 0.031 \\
 16.0  &  0.042 $\pm$ 0.021 &   68.0  &  0.262 $\pm$ 0.032 \\
 18.0  &  0.145 $\pm$ 0.022 &   70.0  &  0.277 $\pm$ 0.031 \\
 20.0  &  0.248 $\pm$ 0.026 &   72.0  &  0.192 $\pm$ 0.034 \\
 22.0  &  0.283 $\pm$ 0.028 &   74.0  &  0.213 $\pm$ 0.033 \\
 24.0  &  0.397 $\pm$ 0.035 &   76.0  &  0.102 $\pm$ 0.039 \\
 26.0  &  0.385 $\pm$ 0.018 &   78.0  & $-$0.055 $\pm$ 0.055 \\
 28.0  &  0.338 $\pm$ 0.013 &   80.0  & $-$0.160 $\pm$ 0.054 \\
 30.0  &  0.294 $\pm$ 0.014 &   82.0  & $-$0.221 $\pm$ 0.051 \\
 32.0  &  0.211 $\pm$ 0.014 &   84.0  & $-$0.217 $\pm$ 0.037 \\
 34.0  &  0.145 $\pm$ 0.015 &   86.0  & $-$0.272 $\pm$ 0.063 \\
 36.0  &  0.070 $\pm$ 0.018 &   88.0  & $-$0.440 $\pm$ 0.066 \\
 38.0  & $-$0.020 $\pm$ 0.017 &   90.0  & $-$0.339 $\pm$ 0.044 \\
 40.0  & $-$0.025 $\pm$ 0.027 &   92.0  & $-$0.441 $\pm$ 0.038 \\
 42.0  & $-$0.102 $\pm$ 0.035 &   94.0  & $-$0.514 $\pm$ 0.036 \\
 44.0  & $-$0.212 $\pm$ 0.041 &   96.0  & $-$0.457 $\pm$ 0.079 \\
 46.0  & $-$0.169 $\pm$ 0.044 &   98.0  & $-$0.335 $\pm$ 0.066 \\
 48.0  & $-$0.189 $\pm$ 0.034 &  100.0  & $-$0.389 $\pm$ 0.083 \\
 50.0  & $-$0.156 $\pm$ 0.041 &  104.0  & $-$0.372 $\pm$ 0.073 \\  
 52.0  & $-$0.132 $\pm$ 0.045 &         &                    \\
\noalign{\smallskip}\hline
\end{tabular}
}
\end{minipage}
\label{table_ayy}
\end{table}

\begin{figure}
\resizebox{0.48\textwidth}{!}{
  \includegraphics{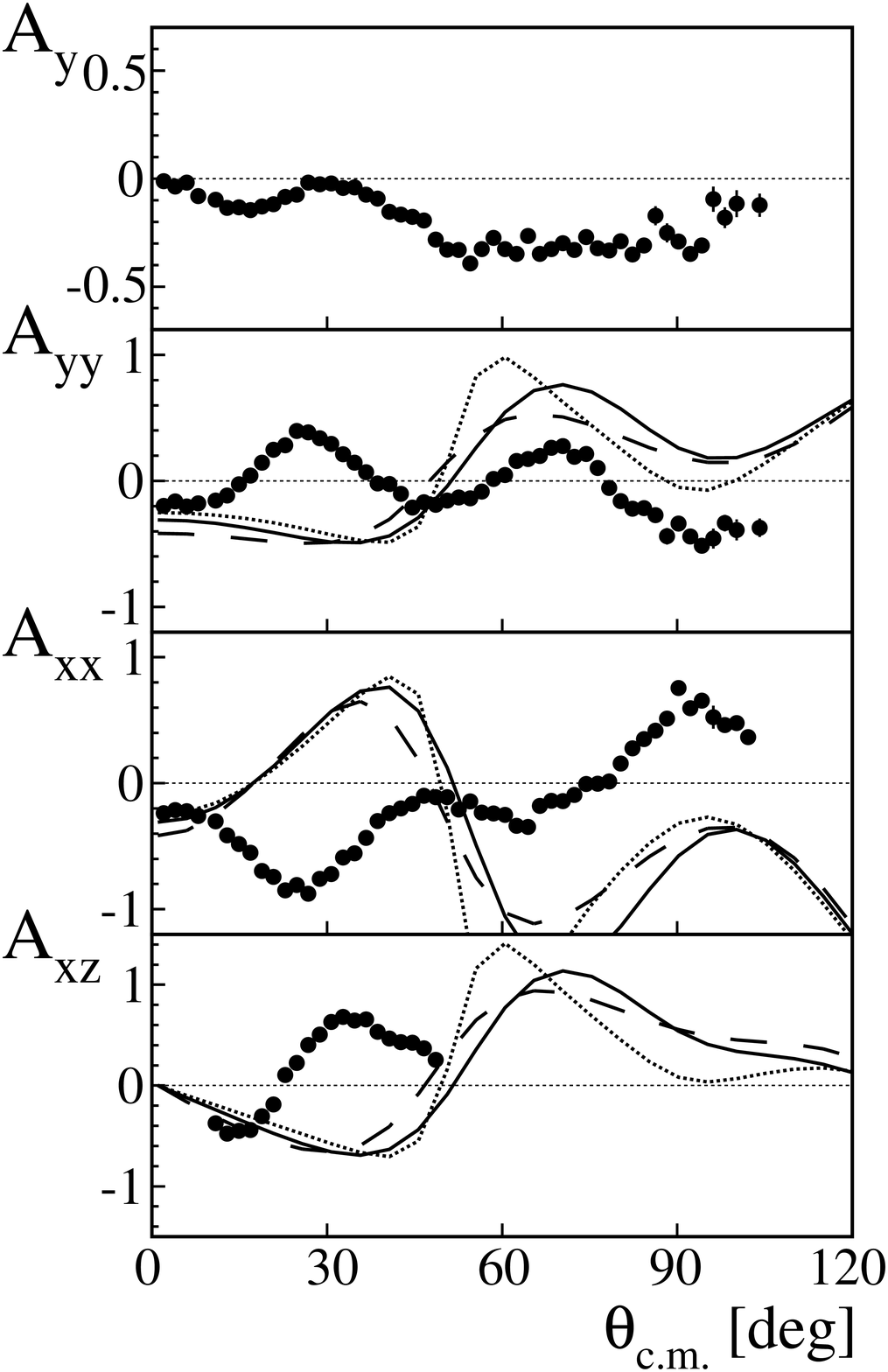}
}
\caption{ 
The results on the vector $A_y$ and tensor $A_{yy}$, $A_{xx}$ and $A_{xz}$ analyzing powers  
	at energy $T_d$=270 MeV as a function of the  angle in the c.m. 
The solid, dashed and dotted curves are the results of the non-relativistic ONE calculations
	\cite{lad,r308n}  using Urbana \cite{urbana}, Paris \cite{laget5} and RSC \cite{rsc} 
	 ${\rm ^3He}$ wave functions, respectively.
Paris deuteron wave function \cite{paris} was used to describe the deuteron structure.
}
\label{ayayyaxxaxz}
\end{figure}

The angular dependence of the vector $A_y$ and tensor $A_{yy}$, $A_{xx}$ and $A_{xz}$ analyzing powers
	at the energy $T_d$=270 MeV are presented in Fig.\ref{ayayyaxxaxz}.
The errors of the analyzing powers  include both the statistical and systematic errors due to the uncertainty in the beam
	polarization. 
One can see  strong variations of the analyzing powers as a function of the angle in the c.m.

The negative sign of $A_{yy}$ and $A_{xx}$ values at small scattering angles is in a striking contrast to
	the positive $A_{yy}$ and $A_{xx}$ for the ${\rm dp\to pd}$  \cite{kimiko1,punjabi}
	or $d{\rm ^3He}\to p{\rm ^4He}$ \cite{uesaka} reactions where the
	deuteron structure is relevant. 
The negative tensor analyzing powers 	can be understood in terms 
	of the ratio of  the $D$ and $S$ wave component of the ${\rm ^3He}$ wave function 
	by means of ONE calculations.

\begin{figure}
\resizebox{0.48\textwidth}{!}{
\includegraphics{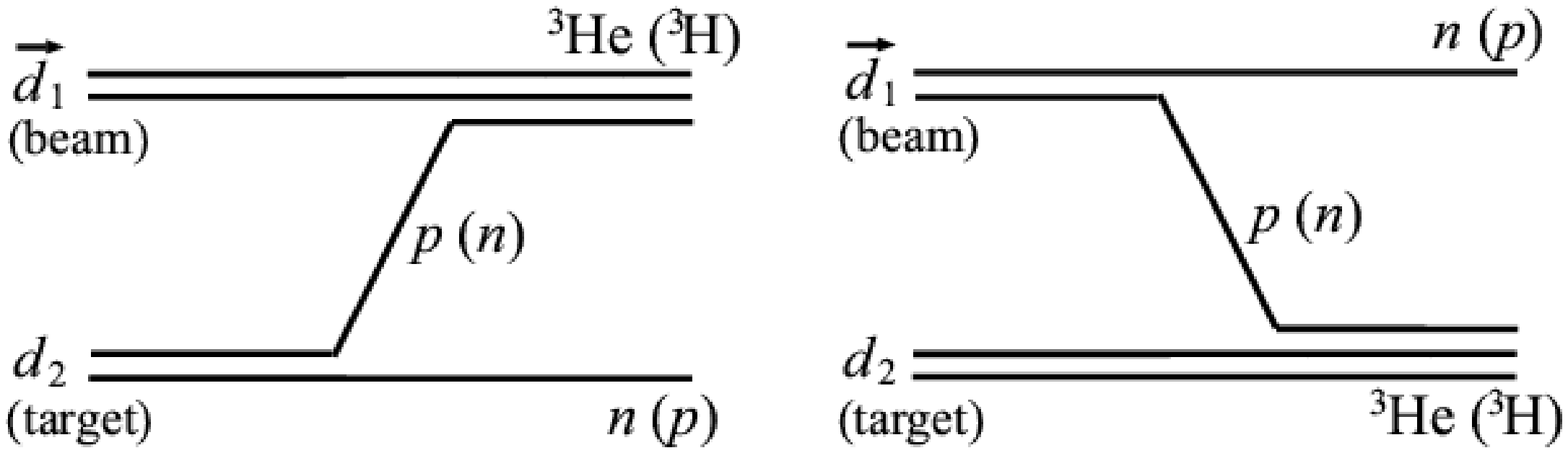}
}
\caption{
ONE diagrams for the $dd\to{\rm ^3He}n({\rm ^3H}p)$ reaction.
}
\label{diagram}
\end{figure}

Within the framework of ONE approximation the $dd\to{\rm ^3He}n$ process 
	can be described by a sum of 2 diagrams 
	(see Fig.\ref{diagram}) required by the symmetry of the initial state of the reaction. 
If the ${\rm ^3He}$ is scattered at forward angles 
	the contribution of the second diagram becomes negligible 
	due to a large relative momentum between the nucleons in the deuteron.
Consequently, only the first diagram gives the contribution 
	to the cross section and polarization observables. 
It has been found \cite{l2004} that the tensor analyzing powers due to
	polarization of the  deuteron beam are sensitive to the ratio of the $D$ and $S$ 
	wave component of the ${\rm ^3He}$ and deuteron wave function, 
	when ${\rm ^3He}$ is emitted in the forward and backward directions in the c.m., respectively.

The solid, dashed, and dotted curves in Fig.\ref{ayayyaxxaxz} 
	are the results of non-relativistic  ONE calculations \cite{lad,r308n} using Urbana \cite{urbana}, Paris \cite{laget5} 
	and RSC \cite{rsc} (with the parametrization from \cite{uzikov}) wave functions of ${\rm ^3He}$.
The Paris parametrization \cite{paris}  was applied for the deuteron wave function.

The negative sign of the tensor analyzing powers $A_{yy}$ and $A_{xx}$ at small scattering angles reflects
	the  positive sign of the ratio of the $D/S$ wave component of the ${\rm ^3He}$ wave function 
	 in the momentum space. 
This behaviour of the data is consistent with the $D$-state admixture in the ${\rm ^3He}$ predicted
	in several theoretical works  \cite{theory,santos}.
However, the trend of the tensor analyzing powers $A_{yy}$ and  $A_{xx}$ at the 
	angles below 15$^\circ$ in the c.m. is opposite to the ONE calculations.

\begin{table}
\begin{minipage}[h]{75mm}
{
\caption{The angular dependence of the tensor  analyzing power $A_{xx}$ in the 
$dd\to {\rm ^3He}n$ reaction at 270 MeV.}
}  
{\centering 
\begin{tabular}{llll}
\hline\noalign{\smallskip}
$\theta_{c.m.}$ &  ~~$A_{xx}\pm\Delta A_{xx}$    & $\theta_{c.m.}$ &  ~~$A_{xx}\pm\Delta A_{xx}$       \\
\noalign{\smallskip}\hline\noalign{\smallskip} 
  1.0  &  $-$0.237 $\pm$ 0.028 & 54.0  &  $-$0.144 $\pm$ 0.049  \\
  3.0  &  $-$0.210 $\pm$ 0.017 & 56.0  &  $-$0.233 $\pm$ 0.058  \\
  5.0  &  $-$0.221 $\pm$ 0.023 & 58.0  &  $-$0.241 $\pm$ 0.047  \\
  7.0  &  $-$0.260 $\pm$ 0.025 & 60.0  &  $-$0.252 $\pm$ 0.062  \\
 10.0  &  $-$0.302 $\pm$ 0.041 & 62.0  &  $-$0.338 $\pm$ 0.042  \\
 12.0  &  $-$0.413 $\pm$ 0.041 & 64.0  &  $-$0.347 $\pm$ 0.043  \\
 14.0  &  $-$0.482 $\pm$ 0.045 & 66.0  &  $-$0.179 $\pm$ 0.046  \\ 
 16.0  &  $-$0.551 $\pm$ 0.007 & 68.0  &  $-$0.139 $\pm$ 0.033  \\
 18.0  &  $-$0.697 $\pm$ 0.034 & 70.0  &  $-$0.141 $\pm$ 0.026  \\
 20.0  &  $-$0.744 $\pm$ 0.038 & 72.0  &  $-$0.092 $\pm$ 0.027  \\
 22.0  &  $-$0.851 $\pm$ 0.038 & 74.0  &  $-$0.006 $\pm$ 0.038  \\
 24.0  &  $-$0.808 $\pm$ 0.041 & 76.0  &  $-$0.003 $\pm$ 0.053  \\
 26.0  &  $-$0.877 $\pm$ 0.039 & 78.0  &   0.013 $\pm$ 0.048  \\
 28.0  &  $-$0.759 $\pm$ 0.036 & 80.0  &   0.156 $\pm$ 0.054  \\
 30.0  &  $-$0.722 $\pm$ 0.038 & 82.0  &   0.277 $\pm$ 0.043  \\
 32.0  &  $-$0.589 $\pm$ 0.037 & 84.0  &   0.351 $\pm$ 0.038  \\
 34.0  &  $-$0.556 $\pm$ 0.037 & 86.0  &   0.417 $\pm$ 0.042  \\
 36.0  &  $-$0.435 $\pm$ 0.039 & 88.0  &   0.512 $\pm$ 0.046  \\
 38.0  &  $-$0.301 $\pm$ 0.027 & 90.0  &   0.755 $\pm$ 0.056  \\
 40.0  &  $-$0.239 $\pm$ 0.034 & 92.0  &   0.595 $\pm$ 0.029  \\
 42.0  &  $-$0.199 $\pm$ 0.036 & 94.0  &   0.656 $\pm$ 0.043  \\
 44.0  &  $-$0.165 $\pm$ 0.036 & 96.0  &   0.524 $\pm$ 0.091  \\
 46.0  &  $-$0.099 $\pm$ 0.040 & 98.0  &   0.461 $\pm$ 0.052  \\
 48.0  &  $-$0.110 $\pm$ 0.044 & 100.0 &   0.478 $\pm$ 0.039  \\
 50.0  &  $-$0.112 $\pm$ 0.036 & 102.0 &   0.365 $\pm$ 0.051  \\
 52.0  &  $-$0.209 $\pm$ 0.052 &       &                      \\ 
\noalign{\smallskip}\hline
\end{tabular}
}
\end{minipage}
\label{table_axx}
\end{table}

\begin{table}
\begin{minipage}[h]{75mm}
{
\caption{The angular dependence of the tensor  analyzing power $A_{xz}$ in the 
$dd\to {\rm ^3He}n$ reaction at 270 MeV.}
}  
{\centering 
\begin{tabular}{llll}
\hline\noalign{\smallskip}
$\theta_{c.m.}$ &  ~~$A_{xz}\pm\Delta A_{xz}$    & $\theta_{c.m.}$ &  ~~$A_{xz}\pm\Delta A_{xz}$       \\
\noalign{\smallskip}\hline\noalign{\smallskip}
 10.0  & $-$0.374 $\pm$ 0.024 & 30.0  &  0.630 $\pm$ 0.025 \\
 12.0  & $-$0.478 $\pm$ 0.027 & 32.0  &  0.682 $\pm$ 0.027 \\
 14.0  & $-$0.449 $\pm$ 0.031 & 34.0  &  0.644 $\pm$ 0.030 \\
 16.0  & $-$0.442 $\pm$ 0.032 & 36.0  &  0.657 $\pm$ 0.035 \\
 18.0  & $-$0.305 $\pm$ 0.040 & 38.0  &  0.532 $\pm$ 0.024 \\
 20.0  & $-$0.188 $\pm$ 0.034 & 40.0  &  0.465 $\pm$ 0.032 \\
 22.0  &  0.103 $\pm$ 0.037 & 42.0  &  0.430 $\pm$ 0.035 \\
 24.0  &  0.224 $\pm$ 0.041 & 44.0  &  0.424 $\pm$ 0.036 \\
 26.0  &  0.401 $\pm$ 0.046 & 46.0  &  0.370 $\pm$ 0.044 \\
 28.0  &  0.505 $\pm$ 0.051 & 48.0  &  0.256 $\pm$ 0.059 \\
\noalign{\smallskip}\hline
\end{tabular}
}
\end{minipage}
\label{table_axz}
\end{table}

The strong disagreement  of the experimental  data from the  non-relativistic  ONE calculations
	\cite{lad,r308n} is observed at  angles larger than 15$^\circ$ in the c.m. 
 The discrepancy between the data and the calculations shown in Fig.\ref{ayayyaxxaxz} can be explained
	by the reaction mechanism which differs from ONE and/or by the non-adequate description 
	of the short-range  ${\rm ^3He}$ spin structure. 
One of the additional mechanisms can be  the $\Delta$-isobar excitation. 
This mechanism 	has been taken into account  phenomenologically
	to describe the $T_{20}$ data in the $d{\rm ^3He}$- backward elastic scattering \cite{dhe3}.
The microscopic calculation by Laget et al.\cite{laget5} has shown 
	that the coherent sum of ONE and the $\Delta$-isobar excitation reasonably reproduces
	 the cross section for the $dd\to{\rm ^3He}n$ reaction at GeV energies.
The calculation predicts that the $\Delta$-isobar
	 contribution to the cross section is $10\%$ at most in the energy region lower than 300~MeV.
It is a dominating  contribution to the ONE process. 
On the other hand, our data on the vector analyzing power $A_y$ have values of $\sim$$-$0.35
	at the angles larger than  50$^\circ$, while ONE predicts  vanishing vector analyzing powers. 
Thus our data have clearly indicated that the processes which are not included 
	in the calculations in ref.\cite{l2004} are important in this angular region.

\begin{figure}
\resizebox{0.48\textwidth}{!}{
  \includegraphics{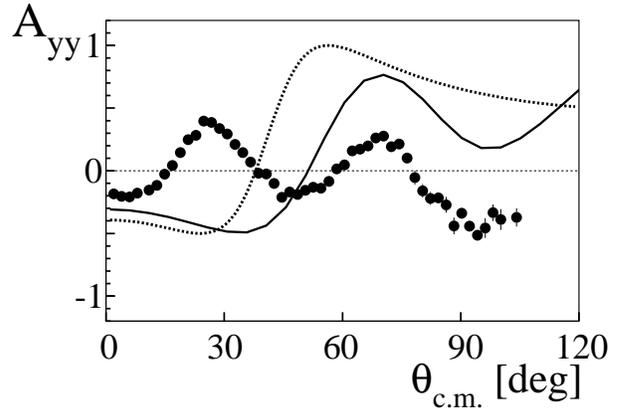}
}
\caption{ 
The results on  the tensor analyzing power $A_{yy}$  at energy $T_d$=270 MeV as a function of the 
${\rm ^3He}$ scattering angle in the c.m. 
The solid and dashed curves are the results of ONE calculations
using non-relativistic and relativistic Urbana \cite{urbana}  ${\rm ^3He}$ wave functions, respectively.
}
\label{rel}
\end{figure}

The analysis of the experimental data on the cross sections of the
	$dp\to pd$ and $dd\to{\rm ^3He}n$ reactions \cite{uzikov_jetp} has shown that
	non-nucleonic degrees of freedom can occur already at $T_d\sim$500~MeV. 
The large angles in the c.m. in the present experiment correspond to the short internucleonic distances where 
	the manifestation of non-nucleonic degrees of freedom is possible. 
On the other hand, the  discrepancy between the data on the tensor analyzing powers and 
	ONE calculations \cite{lad,r308n} can be caused by the relativistic effects. 
In Fig.\ref{rel} the tensor analyzing power $A_{yy}$ in the $dd\to {\rm ^3He}n$ reaction
	at 270 MeV is compared with the results of  ONE calculations 
	 using the relativistic and non-relativistic Urbana ${\rm ^3He}$ wave function \cite{urbana}
	shown by the dashed and solid lines, respectively.
Relativity in ${\rm ^3He}$ wave function is taken into account by the minimal
	relativization scheme \cite{lcd}, where a non-relativistic argument of
	the wave function is replaced by the light-cone variable $k$ 
	(with the corresponding renormalization of the wave function) \cite{lcd1}:
\begin{eqnarray}
&&k^2 = \frac{ [\alpha^2 m_{{\it d}T}^2 - (1-\alpha)^2 m_{{\it p}T}^2 ]^2}
{4\alpha(1-\alpha)[\alpha m_{{\it d}T}^2 +(1-\alpha)m_{{\it p}T}^2 ]}
+p_T^2,\nonumber\\
&&m_{{\it d}T}^2 = m_{\it d}^2+ p_T^2,
~~~~~~~~m_{{\it p}T}^2 = m_{\it p}^2+ p_T^2 \mbox{ ,}
\end{eqnarray}
	where $m_{{\it d}}$ and $m_{{\it p}}$ are the deuteron and proton masses, 
	$m_{{\it dT}}$ and $m_{{\it pT}}$ are the deuteron and proton transversal masses, 
	$\alpha$ is the longitudinal momentum fraction taken away by the deuteron in the infinite momentum frame, 
	and $p_T$ is the transverse momentum. 

One can see that  the use of the relativistic ${\rm ^3He}$ wave function \cite{lcd} does not
	allow one to reproduce $A_{yy}$ data. 
The  structure of ${\rm ^3He}$ can be more 
	complicated and  depends on more than one variable
	as in the case of the deuteron where the strong dependence of the spin structure on
	two variables was observed \cite{ayy}.
On the other hand, the relativistic effects for the both reaction mechanisms  
	and ${\rm ^3He}$ structure should be treated in the consistent way. 
For instance, if one takes the relativistic kinematics, boost effects and  Wigner spin rotations,
	it finally leads to rather small effects in the cross section and polarization
	observables  in $Nd$- elastic scattering \cite{witala}.

The observed negative sign of the tensor analyzing powers
	$A_{yy}$, $A_{xx}$ and $A_{xz}$ at small angles has demonstrated
	the sensitivity to the ratio of the $D$ and $S$ wave 
	component of the ${\rm ^3He}$
	wave function.
However, the deviation of the experimental data from the
	ONE calculations at large angles can be due to not only the nonadequate
	description of the short range ${\rm ^3He}$ spin structure, 
	but also to the influence of the mechanisms additional to ONE.
The measurements of the polarization observables in the $dd\to{\rm ^3He}n$
 	process can provide independent information on the ${\rm ^3He}$ 
	spin structure with respect to the data 
	for the ${\rm ^3He}(p, pN)$ \cite{mil7} and $dp\to{^3He}\gamma$ 
	\cite{meh07,sag03} reactions, where the rescattering
	and meson-exchange current effects play an important role 
	and mask the structure of ${\rm ^3He}$.

\section{Conclusions}
\label{conclusions}

\begin{sloppypar}

The high precision data on the  $A_{yy}$, $A_{xx}$, $A_{xz}$ and $A_y$
	analyzing powers in the $dd\to{\rm ^3He}n$ reaction at the energy 270~MeV 
	have been obtained.

The ONE calculations  using   the standard  ${\rm ^3He}$ wave functions,
	non-relativistic and relativistic in the minimal scheme \cite{lcd},  have
	described qualitatively the data on the tensor analyzing powers  $A_{yy}$,
	$A_{xx}$ and $A_{xz}$ at small angles.
But they have failed  to reproduce  strong variations of the tensor
	analyzing powers  as a function of the angle in the c.m. 
According to the
	calculations \cite{laget5} the $dd\to{\rm ^3He}n$
	reaction is dominated by ONE at these energies. 
The	$\Delta$-isobar contribution is less than $10\%$ at energies
	lower than 300 MeV \cite{laget5}. 
The deviation of the experimental data from
	the ONE calculations can be explained  by the
	nonadequate description of the short range spin ${\rm ^3He}$
	structure, for instance, manifestation of non-nucleonic degrees of freedom
	within the theoretical model considered here.
On the other hand, it is possible that not considered 
	above  reaction mechanisms can affect   the polarization data.

The observed features and high precision of the obtained data from
	the present experiment put serious constraints on the models describing
	the  ${\rm ^3He}$ structure.
However, additional measurements of the  polarization
	observables  in the $dd\to{\rm ^3He}n$ reaction at different energies 
	as well as further theoretical calculations are required 
	to improve the description  of the obtained data.
In this respect, our data are important to
	study the $dd\to{\rm ^3He}n$ reaction as a probe to explore
	the short range spin structure of three nucleon bound state.

\vskip 5mm
The authors express their thanks to the staff of RARF having provided
	excellent conditions for the R308n experiment.
Deep recognition and appreciation is expressed to  H.Kumasaka, R.Suzuki and R.Taki for their help
	during the experiment.
The Russian part of collaboration thanks the RIKEN Directorate for kind 
	hospitality while their stay in Japan.
The investigation has been  partly supported 
by the Grant-in-Aid for Scientific Research (Grant No. 14740151) of 
the Ministry of Education, Culture, Sports, Science, and Technology of Japan;
by the Russian Foundation for 
Fundamental Research (Grant No. 07-02-00102-a) and by the Grant Agency
for Science at the Ministry of Education of the Slovak Republic 
(Grant  No. 1/4010/07).

\end{sloppypar}

\end{document}